\documentclass{PoS}
\usepackage[utf8]{inputenc}
\usepackage{amsmath,xspace}
\usepackage{subfigure}
\newcommand\url[1]{\href{#1}{\tt #1}}

\title{NLO matching for $\boldsymbol{t\bar{t}b\bar{b}}$ production with massive $\boldsymbol{b}$ quarks}

\ShortTitle{NLO matching for $t\bar{t}b\bar{b}$ production with massive $b$ quarks}

\author{\speaker{Tom\'a\v{s} Je\v{z}o}\thanks{Work supported by the Swiss National Science Foundation~(SNF) under
contracts BSCGI0-157722 and CRSII2-160814.}\\
        Physics Institute, Universit\"at Z\"urich, Z\"urich, Switzerland\\
        E-mail: \email{tomas.jezo@physik.uzh.ch}}

\newcommand{\Powhel}{{\rmfamily\scshape Powhel}\xspace}
\newcommand{\Powheg}{{\rmfamily\scshape Powheg}\xspace}
\newcommand{\PowhegBox}{{\rmfamily\scshape Powheg-Box}\xspace}
\newcommand{\PowhegBoxRes}{{\rmfamily\scshape Powheg-Box-Res}\xspace}

\newcommand{\PowhegPythia}{{\rmfamily\scshape Powheg+Pythia8}\xspace}

\newcommand{\Sherpa}{{\rmfamily\scshape Sherpa}\xspace}
\newcommand{\SherpaOpenLoops}{{\rmfamily\scshape Sherpa+OpenLoops}\xspace}
\newcommand{\OpenLoops}{{\rmfamily\scshape OpenLoops}\xspace}
\newcommand{\MadgraphaMC}{{\rmfamily\scshape Madgraph5aMC@NLO}\xspace}
\newcommand{\Pythia}{{\rmfamily\scshape Pythia8}\xspace}

\newcommand{\GeV}{\text{GeV}\xspace}

\newcommand{\rB}{\mathrm{B}}
\newcommand{\rad}{\mathrm{rad}}
\newcommand{\rT}{\mathrm{T}}

\newcommand{\ttbar}{\ensuremath{t \bar t}\xspace}
\newcommand{\bbbar}{\ensuremath{b \bar b}\xspace}
\newcommand\ttbb{\ensuremath{t \bar t b\bar b}\xspace}
\newcommand{\Nb}{N_b}

\def\beq{\begin{equation}}
\def\beqn{\begin{eqnarray}}
\def\eeq{\end{equation}}
\def\eeqn{\end{eqnarray}}

\def\({\left(} 
\def\){\right)} 
 
\newcommand     \MSB            {\ifmmode {\overline{\rm MS}} \else
                                 $\overline{\rm MS}$\fi}

\newcommand\as{\alpha_{\rm S}}

\newcount\minutes 
\newcount\scratch 
\def\timestamp{%
\scratch=\time 
\divide\scratch by 60 
\edef\hours{\the\scratch} 
\multiply\scratch by 60 
\minutes=\time 
\advance\minutes by -\scratch 
---$\,$\hours:\null 
\ifnum\minutes< 10 0\fi 
\the\minutes}

\def\reffi#1{\mbox{Fig.~\ref{#1}}}

\def\citere#1{\mbox{Ref.~\cite{#1}}}

\newcommand{\sing}{\mathrm{s}}
\newcommand{\fin}{\mathrm{f}}

\newcommand{\hdamp}{\ensuremath{h_{\mathrm{damp}}}\xspace}

\newcommand{\hbzd}{\ensuremath{h_{\mathrm{bzd}}}\xspace}
\newcommand{\calR}{\mathcal{R}}
\newcommand{\calK}{\mathcal{K}}

\newcommand{\ie}{i.e.~}

\abstract{
Measurements of $\ttbar H$ production in the $H\to \bbbar$ channel depend in a critical way on the theoretical uncertainty associated with the irreducible QCD $\ttbar+b$-jet background.
We introduce a new $pp\to \ttbb$ \Powheg generator in the 4F scheme based on \PowhegBoxRes and on \OpenLoops for fast evaluation of the scattering amplitudes.
We present predictions and uncertainties for $\ttbar+b$-jet observables at the 13\,TeV LHC.
We also consider theoretical uncertainties related to the \Powheg matching method and to the parton shower (PS) modelling, with emphasis on $g\to \bbbar$ splittings.
In general, matching and shower uncertainties turn out to be remarkably small. 
This is confirmed by a consistent comparison against \SherpaOpenLoops.
}

\FullConference{XXVI International Workshop on Deep-Inelastic Scattering and Related Subjects (DIS2018)\\
		16-20 April 2018\\
		Kobe, Japan}

\begin{document}

\section{Introduction}

At the Large Hadron Collider (LHC), searches for $\ttbar H$ production in the $H\to \bbbar$ channel are plagued by a large QCD background, which is dominated by $\ttbb$ production.
The availability of precise theoretical predictions for this multi-particle background process is of crucial importance for the sensitivity of $\ttbar H(\bbbar)$ analyses.
The process $pp\to \ttbb$ is also very interesting on its own, as it provides a unique laboratory to explore the QCD dynamics of heavy-quark production and to test state-of-the-art Monte Carlo predictions in a nontrivial multi-scale environment.

As a result of its $\as^4$ dependence, the leading-order (LO) $\ttbb$ cross section is highly sensitive to variations of the renormalisation scale.
The uncertainty corresponding to standard factor-two scale variations amounts to 70--80\%, and the inclusion of next-to-leading order (NLO) QCD corrections~\cite{Bredenstein:2009aj,Bevilacqua:2009zn,Bredenstein:2010rs}, where the scale dependence reduces down to 20--30\%, is mandatory.
In~\citere{Cascioli:2013era} it was pointed out that matching and shower effects also play an unexpectedly important role in $\ttbar+b$-jet production.
This is due to the fact that two hard $b$ jets can arise from two hard jets each involving a collinear $g\to \bbbar$ splitting.
In simulations of $pp\to \ttbb$ at NLO and matched to PS (NLO+PS), such configurations result from the combination of a $g\to \bbbar$ splitting that is described at NLO accuracy through $\ttbb$ matrix elements together with a second $g\to \bbbar$ splitting generated by the PS.
The impact of this mechanism can have similar magnitude to the $\ttbar H(\bbbar)$ signal, and a thorough understanding of the related matching and shower uncertainties is very important for $\ttbar H$ analyses.

A first assessment of NLO+PS uncertainties was presented in~\citere{deFlorian:2016spz} through a tuned comparison of NLO+PS $\ttbb$ simulations in \Powhel~\cite{TROCSANYI:2014lha,Garzelli:2014aba}, \SherpaOpenLoops~\cite{Cascioli:2013era} and \MadgraphaMC~\cite{Alwall:2014hca}.
On the one hand, this study has revealed significant differences between the two generators based on the MC@NLO matching method, \ie~\Sherpa and \MadgraphaMC. 
On the other hand, in spite of the fact that \SherpaOpenLoops and \Powhel implement different matching methods and different parton showers, the predictions of these two generators turned out to be quite consistent.
However, due to the limitations related to the use of the five-flavour scheme in \Powhel---which have been overcome only very recently with the 4F upgrade of \Powhel~\cite{Bevilacqua:2017cru}---the agreement between \Powhel and \SherpaOpenLoops did not allow any firm conclusions to be drawn in the study of~\citere{deFlorian:2016spz}.

Motivated by the above we present a new \Powheg generator for $pp\to \ttbb$ in the 4F scheme in Ref.~\cite{Jezo:2018yaf}.
At variance with the \Powhel generator of \citere{Bevilacqua:2017cru}, this new \Powheg generator is implemented in the \PowhegBoxRes framework~\cite{Jezo:2015aia} using \OpenLoops, which guarantees a very fast evaluation of the required $2\to 4$ and $2\to 5$ matrix elements, and supports top-quark decays including spin-correlation effects.
Moreover, in order to guarantee a consistent resummation of QCD radiation, the separation of the so-called singular and finite parts in \PowhegBox, is not restricted to initial-state radiation as in~\citere{Bevilacqua:2017cru} but is applied also to final-state radiation.

In these proceedings we briefly review some of the findings of a much more extensive and complete study in Ref.~\cite{Jezo:2018yaf}, that this manuscript derives from.
Specifically, we briefly discuss technical subtleties that arise when matching multi-scale processes like $pp\to \ttbb$ to PS within the \Powheg framework and show a subset of our predictions for $t\bar{t}+b$-jet observables compared to those obtained using \Sherpa. 
We also compare against the predictions of inclusive NLO+PS $t\bar{t}$ generators.

\section{Parton shower matched simulations for $\boldsymbol{t\bar{t}+b}$-jets}

In the following we briefly review some aspects of the \Powheg method~\cite{Nason:2004rx, Frixione:2007vw} pertaining to the separation of radiation into singular and finite parts. 
We pay particular attention to issues related to the multi-scale nature of the process at hand.  
In particular we point out that the treatment of the recoil associated with the real emissions can induce sizeable distortions of the underlying $\ttbb$ cross section. 
This technical inconvenience restricts the domain of applicability of QCD factorisation in a way that can jeopardise the efficiency of event generation and can also lead to unphysical resummation effects.  

The master formula for the description of NLO radiation in the \Powheg approach consists of two contributions due to splitting the real emission into singular and finite parts, respectively denoted by $R_\sing$ to $R_\fin$.
The splitting is achieved via a damping function $F$ that fulfills $F \to 1$ and $F \to 0$ respectively, in the infrared and hard regions of the phase space.
Only the singular contribution is used for the calculation of the Sudakov form factor for the generation of the hardest emission, and thus resummed.
The default functional form of $F$ in \PowhegBox~\cite{Alioli:2010xd,Alioli:2008tz} is $F (\Phi) = \hdamp^2/(\hdamp^2+k_{\rT}^2(\Phi))$.
It smoothly shifts the weight of real radiation from $R_\sing$ to $R_\fin$ when the hardness of the emission, $k_{\rT}$, becomes of the order of the $\hdamp$ parameter or higher.

In addition to the well-known $\hdamp$-dependent damping mechanism, \PowhegBox also implements a theta function of the form $\tilde{F} (\Phi) = \theta\left(\hbzd -R(\Phi)/\calR(\Phi) \right)$\footnote{The $\tilde{F}$ function multiplies $F$ when combined.}~\cite{Alioli:2008gx,Alioli:2010xd}, where $\calR$ corresponds to the infrared (soft and collinear) approximation of the full matrix element. 
By default, the cut-off parameter $\hbzd$ is set equal to 5.
In this way, in the vicinity of IR singularities, where $R/\calR\to 1$, radiative contributions are attributed to $R_\sing$ and resummed.
On the contrary, when the real emission matrix element largely exceeds the IR approximation, the resummation of the full $R/B$ kernel is not well justified, and the corresponding events are attributed to the finite remnant through the theta function.
In the standard \PowhegBox, and in \citere{Bevilacqua:2017cru}, the damping function $\tilde{F}$ is applied only to initial-state radiation.  
However, in the present $\ttbb$ generator we have extended it to all (massless or massive) final-state emitters, that have a singular region associated with it, in order to ensure a consistent resummation of QCD radiation off $b$-quarks.

The requirement $R(\Phi) < \hbzd\,\calK(\Phi_{\mathrm{rad}})\,B(\Phi_{\mathrm{B}})$ was originally introduced in order to avoid possible divergences of $R(\Phi)/B(\Phi_\rB)$ due to the so-called Born zeros, i.e.~phase space regions where $B(\Phi_{\mathrm{B}})\to 0$.
Such divergences are not physical and cancel in the $\bar B/B$ ratio, but they can still lead to dramatic inefficiencies in the event generation.

In the case of $pp\to \ttbb$, such effects can arise from the interplay of soft and collinear enhancements due to NLO light-jet radiation and to the generation of the $\bbbar$ system in regions with $m_{\bbbar}\ll m_{\ttbb}$ and/or $p_{\rT,\bbbar}\ll m_{\ttbb}$.
For example, let us consider a $gg\to \ttbb g$ event with a gluon emission from the initial state.
Its kinematics are generated starting from a $gg\to \ttbb$ Born event through a mapping that creates the required gluon recoil by boosting the final state in the transverse direction, modifying the kinematics of the $\bbbar$ system as well.
Because the \ttbb system has enough energy to absorb the recoil of gluon emissions much harder that the \bbbar system this may lead to a very significant reduction of $p_{\rT,\bbbar}$.
This violates the main assumption that justifies the \Powheg master formula, namely $R_\alpha(\Phi_\alpha)/B(\Phi_B)\sim \calR_\alpha(\Phi_\alpha)/B(\Phi_B)={\cal K}_\alpha(\Phi_\rad)$, which requires a sufficiently hard $\ttbb$ process as compared to the $k_\rT$ of NLO radiation. 
In particular, due to the sensitivity of the Born amplitude to scales of the order ${p_{\rT,\bbbar}}\sim (E_{\bbbar}/E_{\ttbb})\, p_{\rT,j}\ll p_{\rT,j}$, the factorisation condition is not fulfilled.

\section{Setup}
\label{se:setup}

The predictions in this work are obtained using the same choices for the heavy-quark masses, flavour schemes, renormalisation and factorization scales and PDFs as in Ref.~\cite{Jezo:2018yaf}, which also correspond to the setup recommended in~\cite{deFlorian:2016spz}.
The \PowhegBox parameters $\hbzd$ and $\hdamp$, which control the resummation of NLO radiation, have been set to $\hbzd=2$ and $\hdamp=H_\rT/2 = 1/2 \sum_{i=t,\bar t, b, \bar b} E_{\rT,i}$.
To account for the uncertainties associated with this choice we apply the independent variations $\hbzd=2$, $5$, $10$ and $\hdamp={H_\rT}/{4},{H_\rT}/{2},H_\rT,1.5\,m_t$, varying both parameters one at a time.
At LO+PS level we set the shower starting scale equal to $H_{\rT}/2$ and vary it up and down by a factor of two in order to assess the related uncertainty.
At NLO+PS, the shower starting scale is dictated by the kinematics of real emission matrix elements in the \Powheg method.

In order to assess uncertainties due to the parton-shower modelling of $g\to \bbbar$ splittings we vary the parameter {\tt TimeShower:weightGluonToQuark}, which permits one to select out of various forms of the $g\to Q\bar Q$ splitting kernel in \Pythia8, considering the options 2 and 4~\cite{timeshowers}.
In addition to the functional form of the heavy-quark splitting kernel we also vary the scale of $\as$ in the parton shower.
To this end, we set {\tt TimeShower:weightGluonToQuark} to 6 and 8, which corresponds to the options 2 and 4 with $\as(p_T^2)$ replaced by $\as(m_{\bbbar}^2)$ in the heavy-quark splitting kernel.
Moreover, using {\tt TimeShower:renormMultFac}, we vary $\as(p_\rT^2)\to \as(\xi p_\rT^2)$ with prefactors $\xi=0.1,1,10$.
This latter variation is applied to all final-state QCD splittings, i.e.~also splittings of type $g\to gg$, $q\to q g$, etc.

For the reconstruction of jets we use the anti-$k_\rT$~\cite{Cacciari:2008gp} algorithm with $R=0.4$. 
We select jets that fulfil $p_{\rT}>25\,\GeV$ and $|\eta|<2.5$ both for the case of light jets and $b$-jets.
At parton level, we define as $b$-jet a jet that contains at least a $b$-quark, such that jets that contain a $\bbbar$ pair arising from a collinear $g\to \bbbar$ splitting are also tagged as $b$-jets. 

We categorise events according to the number $\Nb$ of $b$-jets that do not arise from top decays and fulfil the acceptance cuts.
For the analysis of cross sections and distributions we consider an inclusive selection with $N_b\ge 1$ and a more exclusive one with $\Nb\ge 2$. 
We refer to them as ttb and ttbb selections, respectively.

\section{Results}
In \reffi{fig:tools1} we compare $\ttbar+b$-jet predictions based on \PowhegPythia and \Sherpa.
This comparison is done both for (N)LO+PS $pp\to \ttbb$ generators in the 4F scheme and for corresponding generators of inclusive  $\ttbar$ production in the 5F scheme\footnote{In the case of \Powheg we use {\tt hvq}~\cite{Frixione:2007nw}.}.  

Input parameters, QCD scales and matching parameters are chosen as coherently as possible across all generators.
In particular, the parameter $\hdamp=H_\rT/2$ in \Powheg is identified with the resummation scale $\mu_Q$ in the SMC@NLO framework of \Sherpa.
Instead, for what concerns the parton showers we simply use standard settings, \ie~we do not try to improve the agreement between generators by tuning the \Pythia and \Sherpa showers.

\newcommand{\toolsplot}[1]{
\renewcommand{\arraystretch}{0}
\begin{minipage}{0.31\textwidth}
\begin{center}
\includegraphics[width=\textwidth,trim={2 9.9mm 2 0},clip]{tools/#1_top}\\
\includegraphics[width=\textwidth,trim={2 9.9mm 2 7.0mm},clip]{tools/#1_center} \\
\includegraphics[width=\textwidth,trim={2 0 2 7.0mm},clip]{tools/#1_bottom}
\end{center}
\end{minipage}
}
\begin{figure}[t!]
\centering
\subfigure[]{
\hspace{-7mm}
\toolsplot{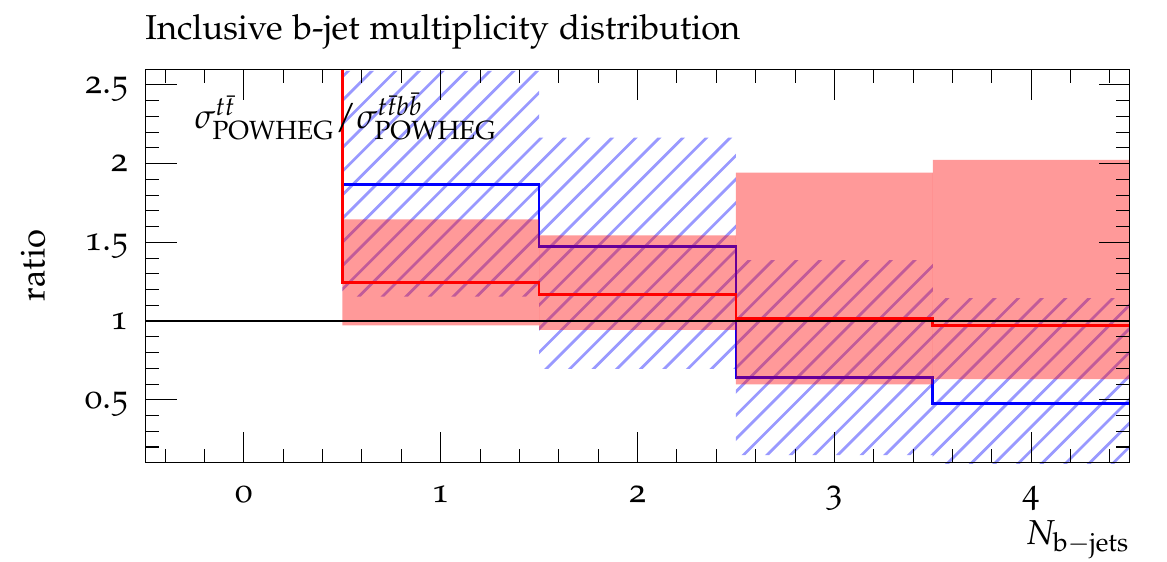}
\label{fig_tools:NBj_XS}}
\subfigure[]{
\hspace{-3mm}
\toolsplot{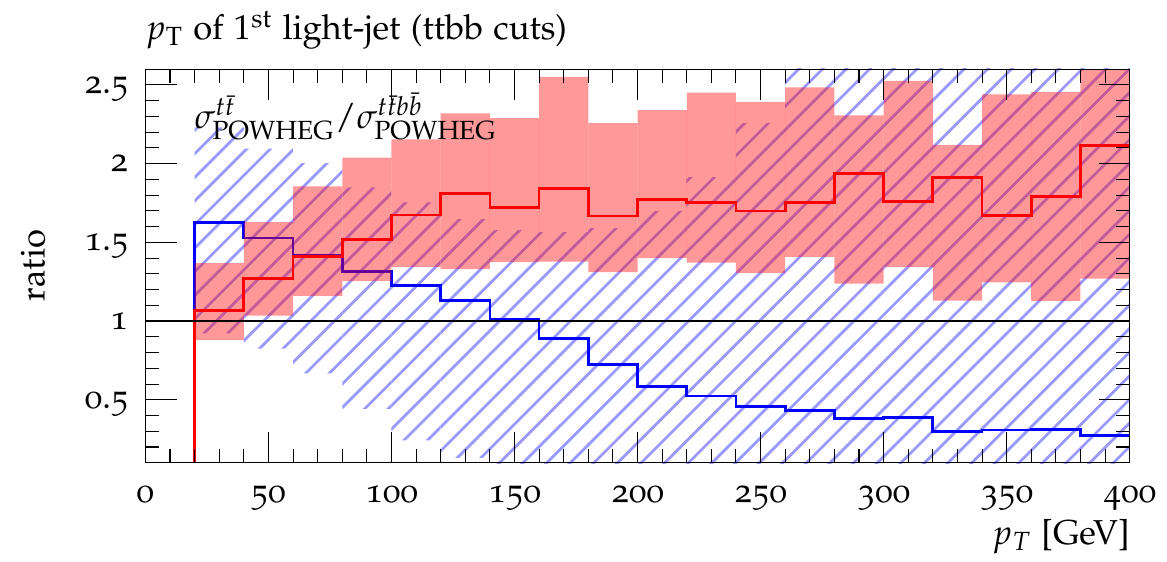}
\label{fig_tools:2_PT_J1}}
\subfigure[]{
\hspace{-3mm}
\toolsplot{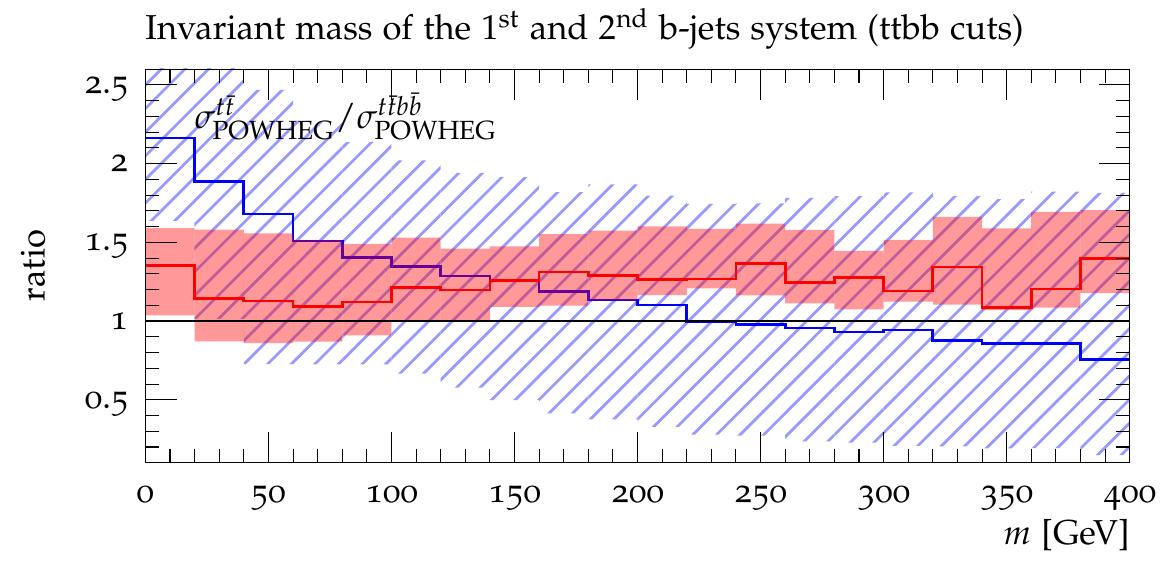}
\label{fig_tools:2_M_B1B2}}
\caption{Predictions for $pp\to \ttbar+b$-jets at $\sqrt{s}=13$\,TeV: distributions in the inclusive number of additional $b$-jets (a), the first light jet (b) with ttbb cuts, and in the invariant mass (c) with ttbb cuts.
The various ratio plots compare $\ttbar+b$-jet observables as described in LO+PS (blue) and NLO+PS (red) simulations based on $pp\to \ttbb$ or $pp\to \ttbar$ matrix elements in \PowhegPythia or \Sherpa.
In the ratios shown in the upper and middle frame \Powheg predictions are normalised to \Sherpa ones for the case of $pp\to \ttbb$ and $pp\to \ttbar$ simulations, respectively.
The third frame displays the ratio of \ttbar to $\ttbb$ \Powheg predictions.
For all ratios the numerator and denominator are evaluated at the same order, and uncertainties are applied only to the numerator.  
They correspond to the combination in quadrature of $\hdamp$ and $\hbzd$ variations with the uncertainties due to the modelling of $g\to \bbbar$ splittings and the choice of $\as$ and the shower starting scale in \Pythia.
}
\label{fig:tools1}
\end{figure}

The ratios in the upper frames show \Powheg $pp\to \ttbb$ predictions normalised to corresponding \Sherpa predictions at LO+PS and NLO+PS accuracy.
The bands describe the combination in quadrature of all matching and shower uncertainties in \PowhegPythia (referred to shower uncertainties in the following), while only nominal \Sherpa predictions are considered in the ratios.
At LO+PS, for observables that are inclusive with respect to jet radiation we find deviations between 10--40\% and comparably large shower uncertainties.
In contrast, in the jet-$p_\rT$ distributions the predictions of \Pythia are far above the ones of \Sherpa, with differences that can reach a factor 2.5 in the tails.
These differences are perfectly consistent with LO+PS shower uncertainties, which are dominated by variations of the \Pythia starting scale.

Moving to NLO+PS reduces the direct dependence on the PS. 
At the same time, differences between the \Powheg and SMC@NLO matching methods come into play.
In practice, at NLO+PS we observe a drastic reduction of shower uncertainties, especially in the light-jet $p_\rT$-distribution.
Also the differences between \Powheg and \Sherpa become very small at NLO+PS. 
The ttb and ttbb cross sections agree at the percent level, and differential $b$-jet observables deviate by more than 5\% only in the tail of the $m_{b_1b_2}$ distribution. 
Even the light-jet spectra in the ttbb phase space deviate by less than 10--20\% up to high $p_\rT$, in spite of the limited formal accuracy (LO+PS) of such observables.
In the light of these results, NLO+PS theoretical uncertainties related to the matching scheme and the PS seem to be well under control in $pp\to \ttbb$ and are clearly subleading as compared to QCD scale uncertainties shown on Figs.~7 and 8 in Ref.~\cite{Jezo:2018yaf}.

In the central frames we compare (N)LO+PS generators of inclusive $\ttbar$ production based on \PowhegPythia and \Sherpa.
In this case, the $g\to \bbbar$ final-state splittings that give rise to $\ttbar+b$-jet signatures are entirely controlled by the PS.
At LO+PS, the parent gluon that splits into $\bbbar$ is also generated by the PS. 
Nevertheless, the ttb and ttbb LO+PS cross sections predicted by \Powheg and \Sherpa deviate by less than 30\%--40\%. 
Instead, as expected, the shapes of $\ttbar+b$-jet observables vary very strongly, and in all considered light-jet and $b$-jet distributions \Pythia results exceed \Sherpa ones by a factor of two and even more. 
This excess is well consistent with the estimated LO+PS shower uncertainties.
At NLO+PS, only $g\to \bbbar$ splittings are controlled by the PS, while the emission of their parent gluon is dictated by LO matrix elements.
Consequently, we observe a drastic reduction of shower uncertainties as compared to LO+PS.
The differences between \Powheg and \Sherpa are also largely reduced at NLO, nevertheless they remain quite significant in various distributions.

To provide a more complete picture of the uncertainties of inclusive $\ttbar$ simulations, in the lower frames we compare \PowhegPythia generators of inclusive $\ttbar$ production and $\ttbb$ production.
Shower uncertainties are shown only for the $\ttbar$ generator.
At LO+PS, the $\ttbar$ generator is strongly sensitive to the modelling of $pp\to \ttbar g$ through initial-state gluon radiation in \Pythia. 
As a result, the $\ttbar$ generator overestimates the ttb and ttbb cross sections by about 90\% and 50\%, respectively.
This excess is strongly sensitive to the shower starting scale, and in the $p_\rT$-distributions it is confined to the regions below 100--200\,GeV, while the tails are strongly suppressed.
Also the $m_{b_1b_2}$ distribution features a strong shape difference as compared to LO+PS $\ttbb$ predictions.

Such differences decrease significantly at NLO+PS.
The ttb and ttbb cross sections predicted by the $\ttbar$ generator overshoot $\ttbb$ results by only 15--20\%, and $b$-jet observables also feature an improved agreement with $\ttbb$ predictions.
Nevertheless, in $b$-jet observables we find quite significant shape differences, especially for the $m_{b_1b_2}$ distribution, and shower uncertainties remain far above the ones of the  $\ttbb$ generator (see upper frame).
As for the light-jet spectra, $\ttbar$ predictions turn out to lie above $\ttbb$ ones by about a factor of two in the tails.

\section{Summary}
Searches for $\ttbar H$ production in the $H\to \bbbar$ channel call for a precise theoretical description of the irreducible QCD $\ttbar+b$-jet background.
We have presented a new $pp\to \ttbb$ \Powheg generator in the 4F scheme based on \OpenLoops and \PowhegBoxRes, the latter requiring a gentle modification in order to overcome subtle technical issues that arise in matching multi-scale processes to PS using the \Powheg method.
At NLO+PS, the matching uncertainties due to \hdamp and \hbzd and shower uncertainties due to the modelling of $g\to \bbbar$ and the choice of $\as$ in \Pythia turn out to be rather small and clearly subleading with respect to QCD scale variations.
We find that with the help of parton-shower tuning, the inclusive NLO+PS $\ttbar$ generators may potentially be amenable to a reasonable description of inclusive $\ttbar+b$-jet observables.
However, it should be clear that NLO+PS $\ttbb$ generators are mandatory in order to achieve an acceptable level of shower systematics.

\section{Acknowledgments}
I am grateful to Silvia Ferrario Ravasio and Katie Whitfield for a careful reading of the manuscript.

\bibliography{paper}

\providecommand{\href}[2]{#2}\begingroup\raggedright\begin{thebibliography}{10}

\bibitem{Bredenstein:2009aj}
A.~Bredenstein, A.~Denner, S.~Dittmaier, and S.~Pozzorini, {\it {NLO QCD
  corrections to $pp \to t \bar{t} b \bar{b} + X$ at the LHC}},  {\em Phys.
  Rev. Lett.} {\bf 103} (2009) 012002,
  [\href{http://arxiv.org/abs/0905.0110}{{\tt arXiv:0905.0110}}].

\bibitem{Bevilacqua:2009zn}
G.~Bevilacqua, M.~Czakon, C.~G. Papadopoulos, R.~Pittau, and M.~Worek, {\it
  {Assault on the NLO Wishlist: $pp \to t \bar{t} b \bar{b}$}},  {\em JHEP}
  {\bf 09} (2009) 109, [\href{http://arxiv.org/abs/0907.4723}{{\tt
  arXiv:0907.4723}}].

\bibitem{Bredenstein:2010rs}
A.~Bredenstein, A.~Denner, S.~Dittmaier, and S.~Pozzorini, {\it {NLO QCD
  Corrections to Top Anti-Top Bottom Anti-Bottom Production at the LHC: 2. full
  hadronic results}},  {\em JHEP} {\bf 03} (2010) 021,
  [\href{http://arxiv.org/abs/1001.4006}{{\tt arXiv:1001.4006}}].

\bibitem{Cascioli:2013era}
F.~Cascioli, P.~Maierh{\"o}fer, N.~Moretti, S.~Pozzorini, and F.~Siegert, {\it
  {NLO matching for $t\bar t b \bar b$ production with massive $b$-quarks}},
  {\em Phys. Lett.} {\bf B734} (2014) 210--214,
  [\href{http://arxiv.org/abs/1309.5912}{{\tt arXiv:1309.5912}}].

\bibitem{deFlorian:2016spz}
{\bf LHC Higgs Cross Section Working Group} Collaboration, D.~de~Florian
  et~al., {\it {Handbook of LHC Higgs Cross Sections: 4. Deciphering the Nature
  of the Higgs Sector}},  \href{http://arxiv.org/abs/1610.07922}{{\tt
  arXiv:1610.07922}}.

\bibitem{TROCSANYI:2014lha}
M.~V. Garzelli, A.~Kardos, and Z.~Trocsanyi, {\it {$t \bar{t} b \bar{b}$
  hadroproduction at NLO accuracy matched with parton shower}},  {\em PoS} {\bf
  EPS-HEP2013} (2013) 253.

\bibitem{Garzelli:2014aba}
M.~V. Garzelli, A.~Kardos, and Z.~Trocsanyi, {\it {Hadroproduction of
  $t\bar{t}b\bar{b}$ final states at LHC: predictions at NLO accuracy matched
  with Parton Shower}},  {\em JHEP} {\bf 03} (2015) 083,
  [\href{http://arxiv.org/abs/1408.0266}{{\tt arXiv:1408.0266}}].

\bibitem{Alwall:2014hca}
J.~Alwall, R.~Frederix, S.~Frixione, V.~Hirschi, F.~Maltoni, et~al., {\it {The
  automated computation of tree-level and next-to-leading order differential
  cross sections, and their matching to parton shower simulations}},  {\em
  JHEP} {\bf 1407} (2014) 079, [\href{http://arxiv.org/abs/1405.0301}{{\tt
  arXiv:1405.0301}}].

\bibitem{Bevilacqua:2017cru}
G.~Bevilacqua, M.~V. Garzelli, and A.~Kardos, {\it {$t\bar{t}b\bar{b}$
  hadroproduction with massive bottom quarks with PowHel}},
  \href{http://arxiv.org/abs/1709.06915}{{\tt arXiv:1709.06915}}.

\bibitem{Jezo:2018yaf}
T.~Ježo, J.~M. Lindert, N.~Moretti, and S.~Pozzorini, {\it {New NLOPS
  predictions for $\boldsymbol{t \bar{t} +b}$ -jet production at the LHC}},
  {\em Eur. Phys. J.} {\bf C78} (2018), no.~6 502,
  [\href{http://arxiv.org/abs/1802.00426}{{\tt arXiv:1802.00426}}].

\bibitem{Jezo:2015aia}
T.~Je\v{z}o and P.~Nason, {\it {On the Treatment of Resonances in
  Next-to-Leading Order Calculations Matched to a Parton Shower}},  {\em JHEP}
  {\bf 12} (2015) 065, [\href{http://arxiv.org/abs/1509.09071}{{\tt
  arXiv:1509.09071}}].

\bibitem{Nason:2004rx}
P.~Nason, {\it {A new method for combining NLO QCD with shower Monte Carlo
  algorithms}},  {\em JHEP} {\bf 11} (2004) 040,
  [\href{http://arxiv.org/abs/hep-ph/0409146}{{\tt hep-ph/0409146}}].

\bibitem{Frixione:2007vw}
S.~Frixione, P.~Nason, and C.~Oleari, {\it {Matching NLO QCD computations with
  Parton Shower simulations: the POWHEG method}},  {\em JHEP} {\bf 11} (2007)
  070, [\href{http://arxiv.org/abs/0709.2092}{{\tt arXiv:0709.2092}}].

\bibitem{Alioli:2010xd}
S.~Alioli, P.~Nason, C.~Oleari, and E.~Re, {\it {A general framework for
  implementing NLO calculations in shower Monte Carlo programs: the POWHEG
  BOX}},  {\em JHEP} {\bf 06} (2010) 043,
  [\href{http://arxiv.org/abs/1002.2581}{{\tt arXiv:1002.2581}}].

\bibitem{Alioli:2008tz}
S.~Alioli, P.~Nason, C.~Oleari, and E.~Re, {\it {NLO Higgs boson production via
  gluon fusion matched with shower in POWHEG}},  {\em JHEP} {\bf 0904} (2009)
  002, [\href{http://arxiv.org/abs/0812.0578}{{\tt arXiv:0812.0578}}].

\bibitem{Alioli:2008gx}
S.~Alioli, P.~Nason, C.~Oleari, and E.~Re, {\it {NLO vector-boson production
  matched with shower in POWHEG}},  {\em JHEP} {\bf 0807} (2008) 060,
  [\href{http://arxiv.org/abs/0805.4802}{{\tt arXiv:0805.4802}}].

\bibitem{timeshowers}
See \url{http://home.thep.lu.se/~torbjorn/pythia82html/TimelikeShowers.html}.

\bibitem{Cacciari:2008gp}
M.~Cacciari, G.~P. Salam, and G.~Soyez, {\it {The anti-$k_T$ jet clustering
  algorithm}},  {\em JHEP} {\bf 04} (2008) 063,
  [\href{http://arxiv.org/abs/0802.1189}{{\tt arXiv:0802.1189}}].

\bibitem{Frixione:2007nw}
S.~Frixione, P.~Nason, and G.~Ridolfi, {\it {A Positive-Weight
  Next-to-Leading-Order Monte Carlo for Heavy Flavour Hadroproduction}},  {\em
  JHEP} {\bf 09} (2007) 126, [\href{http://arxiv.org/abs/0707.3088}{{\tt
  arXiv:0707.3088}}].

\end{thebibliography}\endgroup
\end{document}